\documentclass[aps,twocolumn,groupedaddress,showpacs,amsmath]{revtex4}

\usepackage{epsf,latexsym}
\usepackage{amssymb,amsmath}
\usepackage{times}

\newcommand{\bra}[1]{\langle #1 |}
\newcommand{\ket}[1]{| #1 \rangle}

\newcommand\h{{\cal H}}
\newcommand{\qed}{$\hfill \Box$}

\newcommand\tr{{\mbox{Tr\,}}}

\newcommand{\ignore}[1]{}

\newcommand{\be}{\begin{equation}}
\newcommand{\ee}{\end{equation}}
\newcommand{\ba}{\begin{eqnarray}}
\newcommand{\ea}{\end{eqnarray}}

\def\CC{{\rm\kern.24em \vrule width.04em height1.46ex depth-.07ex
    \kern-.30em C}}
\def\P{{\rm I\kern-.25em P}}

\def\RR{{\rm
         \vrule width.04em height1.58ex depth-.0ex
         \kern-.04em R}}

\def\bbbone{{\mathchoice {\rm 1\mskip-4mu l} {\rm 1\mskip-4mu l}
{\rm 1\mskip-4.5mu l} {\rm 1\mskip-5mu l}}}

\def\bbbc{{\mathchoice {\setbox0=\hbox{$\displaystyle\rm C$}\hbox{\hbox
to0pt{\kern0.4\wd0\vrule height0.9\ht0\hss}\box0}}
{\setbox0=\hbox{$\textstyle\rm C$}\hbox{\hbox
to0pt{\kern0.4\wd0\vrule height0.9\ht0\hss}\box0}}
{\setbox0=\hbox{$\scriptstyle\rm C$}\hbox{\hbox
to0pt{\kern0.4\wd0\vrule height0.9\ht0\hss}\box0}}
{\setbox0=\hbox{$\scriptscriptstyle\rm C$}\hbox{\hbox
to0pt{\kern0.4\wd0\vrule height0.9\ht0\hss}\box0}}}}

\def\bbbz{{\mathchoice {\hbox{$\sf\textstyle Z\kern-0.4em Z$}}
{\hbox{$\sf\textstyle Z\kern-0.4em Z$}}
{\hbox{$\sf\scriptstyle Z\kern-0.3em Z$}}
{\hbox{$\sf\scriptscriptstyle Z\kern-0.2em Z$}}}}

\begin{document}

\title{Quantum entanglement in states generated by bilocal group algebras}

\author{Alioscia Hamma}
\author{Radu Ionicioiu}
\author{Paolo Zanardi}
\affiliation{Quantum Information Group, Institute for Scientific Interchange (ISI), Viale Settimio Severo 65, I-10133 Torino, Italy}

\begin{abstract}
Given a finite group $G$ with a bilocal representation, we investigate the bipartite entanglement in the state constructed from the group algebra of $G$ acting on a separable reference state. We find an upper bound for the von Neumann entropy for a bipartition $(A,B)$ of a quantum system and conditions to saturate it. We show that these states can be interpreted as ground states of generic Hamiltonians or as the physical states in a quantum gauge theory and that under specific conditions their geometric entropy satisfies the entropic area law. If $G$ is a group of spin flips acting on a set of qubits, these states are locally equivalent to 2-colorable (i.e., bipartite) graph states and they include GHZ, cluster states etc. Examples include an application to qudits and a calculation of the $n$-tangle for 2-colorable graph states.
\end{abstract}


\maketitle

\section{Introduction}

Entanglement is certainly the most striking feature of quantum mechanics. Other than for its conceptual importance, entanglement has been in the last years one of the key concepts in quantum information, where it is the main resource required in many protocols of quantum computation and quantum cryptography, e.g., quantum dense coding, Shor's algorithm and teleportation \cite{nielsen_chuang}. Entanglement is also an increasingly important concept in many topics of condensed matter physics, like superconductivity and the fractional quantum Hall effect \cite{zanardi, botero, fqhe}. Moreover, entanglement has recently been used, in the context of quantum phase transitions, as a novel tool to gain insights on the structure of the zero temperature phase diagram of interacting many-body systems \cite{phasetrans}. It also proves to play an important role in understanding certain aspects of quantum field theory \cite{cardy} and spin systems \cite{cirac,keating}; these include 1-dimensional lattice models for $XY$ \cite{latorre,korepin} and Heisenberg models \cite{latorre}.

However, calculating and classifying entanglement for a general physical system is a daunting task. There is no known measure which completely characterizes the entanglement properties of an arbitrary system. If we restrict to bipartite entanglement of a system in a pure state, this task becomes easier. It has been proved that there is an essentially unique entanglement measure, namely the von Neumann entropy of the reduced density matrix of one of the two subsystems, $S= -\tr(\rho_A \log_2 \rho_A)$ \cite{nielsen_chuang, plenio_vedral}. Although this looks conceptually simple, the calculation can be computationally intractable even for simple systems.

On the other hand, in several quantum many-body systems the states of physical interests (e.g., the ground state) are highly symmetric and these symmetries impose additional constraints which can simplify the calculation of entropy. This naturally leads to the topic of this article, namely exploring the entanglement entropy of a bipartite system using a group theoretical framework.

Given a group $G$, possibly non-Abelian, with a bilocal action on a Hilbert space $\h= \h_A\otimes \h_B$, we introduce a class of states, the $G$-states, constructed from the group algebra of $G$ acting on a (separable) reference state. $G$-states emerge naturally as the ground states of generic Hamiltonians. For these states we obtain an upper bound for the entanglement entropy, provided a separability conditions holds for the coefficients. A particular instance of this class, the $G$-homogeneous states, is constructed as an equal superposition of group elements acting on the same reference state $\ket{0}$. As expected, this extra symmetry puts more constraints and is able to provide us an exact formula for the entropy. Remarkably, when $A$ and $B$ represent spatial complementary regions, e.g., in a lattice, we recover the {\em entropic area law}: the entanglement entropy for a bipartition $(A,B)$ depends only on the degrees of freedom localized on the boundary between $A$ and $B$, and not on the bulk ones. This can be viewed as a another manifestation of the Holographic Principle \cite{hp}. The entropic area law has been recovered in several physical systems \cite{srednicki,plenio,wolf}. We examine several examples. If $G$ is a group of spin flips acting on qubits, we show that the corresponding states are Calderbank-Shor-Steane (CSS) states \cite{css} and locally equivalent to a well-known class of stabilizer states, i.e., 2-colorable graph states. These states are important and appear in several physical contexts; examples include the ground state of the Kitaev model of topological quantum computation \cite{kitaev}, error correction codes (the CSS states) and the well known GHZ states. For arbitrary stabilizer states, entanglement has been studied also by Fattal {\em et al.} \cite{fattal}. This group theoretical framework proves to be fruitful also beyond the analysis of bipartite entanglement for qubits. The first extension is to higher dimensional Hilbert spaces, i.e., qudits. The second one applies to multipartite entanglement. Specifically, we show how to calculate the $n$-tangle for a $G$-homogeneous state, where $G$ is a group of qubit spin-flips.

The plan of the article is the following. The framework and the general bounds for the entropy in $G$-states and $G$-homogeneous states are given in Section \ref{gstates}. These are a generalization of our previous results \cite{hiz1,hiz2}. Next we show that we can interpret the $G$-states as the physical states in a gauge theory (Section \ref{gauge}) and we give a geometric interpretation of the entropic bound for $G$-homogeneous states and derive the area law. Section \ref{spinflips} is devoted to $G$-homogeneous states for a group of spin flips and we show that they are locally equivalent to 2-colorable graph states. We also discuss an example for qudits (Section \ref{qudits}) and a calculation of multipartite entanglement (the $n$-tangle) for 2-colorable graph states (Section \ref{ntangle}). We conclude in Section \ref{summary}.

\section{Entanglement in G-states}\label{gstates}

Consider (a unitary representation of) a group $G$ acting on a Hilbert space $\h$. We assume that for a given bipartition $(A,B)$ of the Hilbert space, $\h=\h_A\otimes \h_B$, the action of $G$ on $\h$ is bi-local, i.e., any $g\in G$ is a bi-local operator $g= g_A\otimes g_B$, where $g_{A,B}$ are linear operators acting on the Hilbert space $\h_{A,B}$, respectively. We also assume that it exists a {\em reference product state} $\ket{0}= \ket{0_A} \otimes \ket{0_B}\in \h$. We define the {\em $G$-state} as
\be\label{gstate}
\ket{\Psi_G}:= \sum_{g\in G} \alpha(g) g \ket{0}
\ee
with $\sum_{g,h\in G} \alpha(g) \overline{\alpha} (h) \bra{0} h^{-1}g \ket{0}= 1$.

We point out that $G$-states appear naturally as the ground states of Hamiltonians belonging to the group algebra of $G$, $H \subset \CC (G)$. Let $D=\{g_1,...,g_k\}$ be the set of generators of $G$ with a local structure (i.e., each $g_i$ has a nontrivial action only on a set of neighbouring degrees of freedom). Then we can write any Hamiltonian in the form
\be\label{hamiltonian}
H = \sum_{g_i\in D} \lambda (g_i)g_i
\ee
with $\lambda (g^{-1})= \overline{\lambda}(g)$. Since the ground state of a generic Hamiltonian can be written as $\ket{\psi_0}=\lim_{\beta\rightarrow \infty} Z^{-1}(\beta) e^{-\beta H} \ket{0}$, then $\ket{\psi_0}$ is a sum of all the elements of the group with some coefficients $\lambda '(g):\ket{\psi_0}=\sum_{g\in G}\lambda '(g) g \ket{0}$, hence it is a $G$-state.

If all the coefficients are equal, we call the state a {\em $G$-homogeneous state}
\be
\ket{G}:= N^{-1/2} \sum_{g\in G} g \ket{0}
\label{g_state}
\ee
where the normalization factor $N:= |G| \sum_{g\in G} \bra{0} g\ket{0}\neq 0$ is assumed to be nonzero and $|G|$ is the order of $G$. It is obvious that the state $\ket{G}$ is stabilized by the group $G$, since $h \ket{G}= \ket{G},\ \forall h\in G$. Given a partition $(A,B)$ of the Hilbert space $\h$ we can write the state $\ket{G}$ as
\be
\ket{G} = N^{-1/2} \sum_{g\in G} g_A \ket{0_A}\otimes g_B \ket{0_B}
\ee
where $g_{A,B}$ is the restriction of the operator $g$ to the Hilbert space $\h_{A,B}$. Consider now the two subgroups of $G$ that act exclusively on the subsystems $A$ and respectively $B$:
\begin{eqnarray}
G_A &:=& \{g\in G\ |\ \  g= g_A\otimes \bbbone_B\} \\
G_B &:=& \{g\in G\ |\ \  g= \bbbone_A\otimes g_B\}
\end{eqnarray}
Let their order be $d_{A,B}:= |G_{A,B}|$. Since $G_A$ and $G_B$ are subgroups of the stabilizing group $G$ it is true that $g \ket{G} = \ket{G}$ also for every $g\in G_{A,B}$. As these subgroups are {\em normal}, we can define the quotient group
\be
G_{AB}:=\frac{G}{G_A\times G_B}
\ee
With this notation we have
\be
G= \bigcup_{[h]\in G_{AB}} \{ (g_A\otimes g_B)h\, | \ g_A\otimes\openone\in G_A,\, \openone\otimes g_B\in G_B \}
\label{G_decomp}
\ee

\noindent {\em Proposition 1.} Suppose we have a bipartition $(A,B)$ of a Hilbert space, $\h= \h_A\otimes \h_B$, and that the system is in a $G$-state $\ket{\Psi_G}= \sum_{g\in G} \alpha(g) g\ket{0}$. We assume that for every $g\in G$ there is a representative $h$, with $[h]\in G_{AB}$, such that the coefficients satisfy the separability condition
\be
\alpha(g)\equiv \alpha(g_A\otimes g_B h)= \alpha_A(g_A) \alpha_B(g_B) \beta(h)
\label{separability}
\ee
where again $\ g_A\otimes\openone\in G_A,\, \openone\otimes g_B\in G_B$. Then the von Neumann entropy of the $G$-state corresponding to the bipartition $(A,B)$ is bounded by
\be
S(\ket{\Psi_G})\le -\sum_{[h]\in G_{AB}} |N_A N_B \beta(h)|^2 \log_2 |N_A N_B \beta(h)|^2
\label{S_bound}
\ee
where
\be
N_X^2:= \sum_{g_X\in G_X} |\alpha_X (g_X)|^2, \ \ X= A,B
\label{N_X}
\ee

\noindent {\em Proof.} From Eq.(\ref{G_decomp}) we have
\begin{eqnarray}
\nonumber
\ket{\Psi_G}&=& \sum_{\substack{g_A\otimes g_B\in G_A \times G_B\\ [h]\in G_{AB}}} \alpha_A(g_A) \alpha_B(g_B) \beta(h) (g_A\otimes g_B) h\ket{0}\\
&=& (Q_A\otimes Q_B) \ket{\Psi_G'}
\end{eqnarray}
where $Q_X:= N_X^{-1} \sum_{g_X\in G_X} \alpha_X(g_X) g_X$, with $X=A,B$ and 
\begin{eqnarray}
\nonumber
\ket{\Psi'_G}&=& N_A N_B \sum_{[h]\in G_{AB}} \beta(h) h\ket{0}\\
&=& N_A N_B \sum_{[h]\in G_{AB}} \beta(h) \ket{h_A} \otimes \ket{h_B}
\label{state}
\end{eqnarray}
where $\ket{h_X}:= h_X\ket{0_X}$, $X= A,B$. Since $Q_A\otimes Q_B$ is a bi-local operator, the entanglement satisfies the bound $S(\ket{\Psi_G})\le S(\ket{\Psi_G'})$. The entropy of $\ket{\Psi'_G}$ is maximal when the set $\{ \ket{h_X}\}$, $X= A,B$, is bi-orthogonal (Schmidt decomposition), in which case the entropy is $-\sum_{[h]\in G_{AB}} |N_A N_B \beta(h)|^2 \log_2 |N_A N_B\beta(h)|^2$, hence this proves the bound.  \qed \\

\noindent {\em Observation.} The separability condition (\ref{separability}) seems rather strong. A simple example where this condition is satisfied is $\alpha(g)= \chi^J(g)$, i.e., the coefficients correspond to 1D-characters of the $J$-irrep of $G$.

\noindent {\em Corollary 1.} If for any element $g= g_A\otimes g_B \in G$ we have
\be
g_X \neq \openone_X \Rightarrow \bra{0} g_X \ket{0}=0
\label{zerovev}
\ee
with $X=A,B$, then the entropy saturates its upper bound
\be
S(\ket{\Psi_G})= -\sum_{[h]\in G_{AB}} |N_A N_B \beta(h)|^2 \log_2 |N_A N_B \beta(h)|^2
\label{S1}
\ee
\noindent {\em Proof.}
In order for the entropy $S(\ket{\Psi_G})$ to saturate the previous bound we need to prove first, that the sets $\{ \ket{h_X},\ [h]\in G_{AB} \}$ form a bi-orthogonal basis (hence $S(\ket{\Psi_G'})$ saturates), and second, that $\ket{\Psi_G}$ and $\ket{\Psi_G'}$ have the same entropy. 

Consider now the scalar products $\bra{0} h_X^{-1} h'_X \ket{0}$ with $h_X, h'_X$ such that $[h],[h']\in G_{AB}$ and $X=A, B$. Then $[h]\neq [h'] \Rightarrow h_X^{-1} h'_X= (h^{-1} h')_X \neq \openone$. (If $h_A^{-1} h'_A= \openone$, then $[h^{-1} h']= [\openone]= [h]^{-1}[h'] \Rightarrow [h]=[h']$). From Eq.~(\ref{zerovev}) it follows that the sets $\{ \ket{h_X}\,|\ [h]\in G_{AB}\}$, with $X=A,B$, are a bi-orthogonal basis.

For the second part it is enough to show that the set $\{ \ket{\tilde{h}_X}:= Q_X \ket{h_X},\ X= A,B \}$ is bi-orthogonal. Consider
\begin{eqnarray}
\nonumber
\langle \tilde{h}_X \ket{\tilde{h}_X'}&=& \bra{h_X} Q_X^\dag Q_X \ket{h'_X} \\
\nonumber
&=& N_X^{-2} \sum_{g_X, g'_X \in G_X} \overline{\alpha}_X (g_X) \alpha_X(g'_X) \\
&\times& \bra{0} h_X^{-1} g_X^{-1} g'_X h'_X \ket{0}
\end{eqnarray}
By hypothesis (\ref{zerovev}) the last scalar product is zero unless $h_X^{-1} g_X^{-1} g'_X h'_X= \openone$, hence $h_X= (g_X^{-1} g'_X) h'_X$ and thus $[h_X]= [h_X']$ since they differ by an element of $G_{AB}$. As we can choose the same representative for a given equivalence class, $h'_X= h_X$ and hence $g'_X= g_X$. Taking into account the normalization (\ref{N_X}), we obtain $\langle \tilde{h}_X \ket{\tilde{h}_X'}= \delta_{\tilde{h}_X, \tilde{h}_X'}$. Thus $\ket{\Psi_G}$ and $\ket{\Psi_G'}$ have the same entropy and the thesis follows immediately. \qed

\noindent {\em Corollary 2.} For $G$-homogeneous states the entropy is bounded by
\be
S\le \log_2 |G_{AB}|
\label{S}
\ee
and the bound is saturated if the condition (\ref{zerovev}) holds.

\noindent {\em Corollary 3.} If the group $G$ is a direct product $G= G_A \times G_B$, then $S=0$.

\noindent {\em Proof.} In this case we have $\forall g\in G,\ g= g_A\otimes g_B$ with $g_A\otimes \bbbone_B\in G_A$ and $\bbbone_A\otimes g_B\in G_B$. Then we can write the $G$-state as
\begin{eqnarray}
\nonumber
\ket{\Psi_G} &=& \sum_{g_A\otimes\bbbone_B\in G_A} \alpha_A g_A \ket{0_A}\otimes \sum_{\bbbone\otimes g_B\in G_B} \alpha_B g_B \ket{0_B}\\
&=:& \ket{\chi_A}\otimes\ket{\chi_B}
\end{eqnarray}
As this is a product state with respect to the $(A,B)$ partition, it is obvious that its entanglement is zero. \qed

\section{$G$-states, gauge theories and the entropic area law}\label{gauge}

What is a $G$-homogeneous state? And for which states the bound in Corollary 2 is saturated, i.e., $S=\log_2|G_{AB}|$? The following construction can give us some insight. We assume that $\h$ has a given tensor product structure $\h \simeq \CC^{d_1}\otimes ...\otimes\CC^{d_L}$. Let $\ket{0_{d_i}}$ be a reference vector in $\CC^{d_i}$ and $\ket{0}$ the product state $\ket{0}:= \ket{0_{d_1}} \otimes ...\otimes \ket{0_{d_L}}$. Now let us construct one possible set ${\bf C} := \{ C_1,...,C_L \}$ of linear operators $C_i \in \mathcal L (\CC^{d_i})$, $\forall i=1,..,L$. In the following we take $G$ to be a finite group of linear operators in $\h$ generated by a suitable choice of tensor products of elements in ${\bf C}: G = \langle A_1,...,A_n \rangle$ , where every 
\be
A_k= \bigotimes_{i\in J_k} C_i
\ee
is a local operator on the local Hilbert space $\h_k = \bigotimes_{i\in J_k} \CC^{d_i}$ and $J_k$ is a set of indices. We call the $A_k$ {\em star operators} for reasons that will be clear in the following. A bipartition $(A,B)$ is given by the choice of indices $I_A, I_B$ belonging to the subsystems $A$ and $B$ respectively, such that $I_A\cup I_B=\{1,...,L\}$ and $\h= \h_A\otimes \h_B$, with $\h_A = \bigotimes_{i\in I_A}\CC^{d_i}$, and similarly for $\h_B$. Notice that automatically all the operators in $G$ have a bilocal action on every partition $(A,B)$ of the system. If two sets of indices $J_k, J_{k'}$ have one index in common, $J_k\cap J_{k'}=\{i\}$, we say that the corresponding star operators overlap on the local Hilbert space labeled by $i$. 

Let us now consider $n$ sets of indices $J_1,...,J_n$ defining $n$ star operators such that:
\ba
J_k \subset \bigcup_{r\ne k} J_r,\qquad   k=1,...,n\\
\forall J_k,J_{k'}: J_k\cap J_{k'}=\{i\}\   or \  \emptyset
\ea
We can see a graph (or a lattice) emerging from these $n$ star operators. The vertices of the graph correspond to the star operators, so to $n$ star operators we associate $n$ vertices. The edges of the graph are the local Hilbert spaces  $\CC^{d_i}$ on which the star operators overlap. This justifies their name: a star operator $A_k$ is the product of the operators associated to the edges extruding from a vertex $k$. Star operators obviously have a local structure.

With this construction, we obtain a geometric interpretation of the entanglement in $G$-states. First of all, the entanglement with respect to a partition $(A,B)$ is zero if we can split the set $\{A_k\}$ in two subsets $\{ {\bf A}_K^{(X)} \}$ such that ${\bf A}_K^{(X)}= \bigotimes_{i\in J_K\subset I_X}C_i,\ X= A,B$. So the $A_k$'s in one subset do not overlap with the ones in the other. The two subsets generate together the whole $G$ and since each generates $G_X,\ X=A,B$, respectively, in this case $G$ is a direct product $G= G_A\times G_B$; hence the entanglement is zero from Corollary 3. Geometrically, this means that $A$ and $B$ are not connected in the graph. If the graph is connected then there is no partition for which the entanglement is zero. On the other hand, if no $A_k$'s overlap, the graph is made of all isolated points and the $G$-state is completely disentangled. If we choose the $A_k$'s such that we form a lattice, then there is no way to partition the system such that $G=G_A\times G_B$ and the entanglement is {\em always} different from zero under any possible partition.

We can view the $A_k$'s generating the group $G$ as local gauge symmetries. The vertices of the graph represent points in space. Being $G$ Abelian means that the symmetries act independently at every point. In quantum mechanics we can construct a gauge theory by projecting a Hilbert space $\h$ to a smaller Hilbert space $\h_{phys}$ of the physical states. This is done by requiring that the physical states are annihilated by some operators; for example, the physical states $\ket{\phi_{phys}}$ in quantum electromagnetism are the ones annihilated (at every space-time point) by $\partial_\mu A_\mu \ket{\phi_{phys}}=0$. The Hilbert space $\h$ is the total Hilbert space. The physical states are the states annihilated by the operators $A_k - 1$ \cite{dirac}:
\be
\h_{phys} = \{ \psi \in \h |\ (A_k-1)\psi=0 \}
\ee
The $G$-homogeneous state is obviously a physical state.

How can we characterize all the physical states? Let us find the algebra of the linear operators acting on the physical Hilbert space $\mathcal{L}(\h_{phys})$. What does this algebra look like? Consider the subgroup $\mathcal{W} \subset \mathcal L(\h)$ of linear operators commuting element-wise with $G$ and let $\mathcal{F(W)}$ be its associative algebra. Then consider the ideal $\mathcal I$ generated by the ${A_k-1}$. The algebra acting on the physical Hilbert space is then $\mathcal{L}(\h_{phys})=\mathcal{F(W)/I}$. Then it is immediate to see that
\be\label{Hphys}
\h_{phys}=\mathcal{F(W)/I}\ket{G}= \mbox{span} \{ {\cal W}\ket{G} \} 
\ee
In other words, the orbit of $\mathcal{W}$ through the $G$-homogeneous state $\ket{G}$ is an orthonormal basis in $\h_{phys}$. In general a physical state is not an equal superposition of a group of linear operators, hence we cannot apply Proposition 1 to any quantum physical state of a gauge theory. Nevertheless, consider this particular situation: the system is partitioned such that every element of $\cal W$ is gauge equivalent to some operator acting exclusively on $A$ or $B$. Then, in the hypothesis of $G$ and $\mathcal W$ belonging to the subspace $\mathcal B\subset\mathcal L (\h)$ of linear operators such that $\bra{0} \gamma \ket{0}= 0$ for all $\gamma\in \mathcal B$, the following proposition holds for every physical state in $\h_{phys}$:

\noindent {\em Proposition 2.} Let $(A,B)$ be a partition of $\h$ such that $\cal W$ acts (modulo a gauge transformation $g\in G$) exclusively on either $A$ or $B$, say $w\simeq \bbbone_A\otimes w_B,\; \forall w\in\mathcal{W}$. Then the entanglement entropy of every {\em physical} state $\ket{\psi}\in\h_{phys}$ is equal to $S=\log_2|G_{AB}|$. 

\noindent {\em Proof.} The physical states in the physical Hilbert space $\h_{phys}$ can be written as
\be
\ket{\psi_{phys}} = \sum_{w \in \mathcal{W}} c(w)w \ket{G}
\ee
with the normalization condition $\sum_{w \in \mathcal{W}} |c(w)|^2= 1$. The density matrix of this pure state is
\begin{eqnarray}
\nonumber
\rho^{phys}&=& \sum_{w,w' \in \mathcal{W}} c(w')\overline{c}(w) w'\rho w^{-1} \\
&=& \sum_{w,w' \in \mathcal{W}} c(w')\overline{c}(w^{-1}) w'\rho w
\end{eqnarray}
where $\rho$ is the density matrix of the $G$-homogeneous state, $\rho:=\ket{G}\bra{G}$. Tracing out the $B$ degrees of freedom we obtain the reduced density matrix
\be
\label{pullout1}
\rho_{A}^{phys} =  \sum_{w,w' \in \mathcal {W}} c(w')\overline{c}(w^{-1}) \tr_B (w'\rho w)
\ee
By hypothesis for every element $w \in \mathcal W$ there is a gauge transformation $g\in G$ such that $\tilde{w}:= gw = \bbbone_A \otimes (gw)_B$. Denote $g,g'$ such gauge transformations for the operators $w,w'$ in Eq.(\ref{pullout1}). Then obviously $w\rho w'=wg\rho g'w'= \tilde{w}\rho\tilde{w}'$, since $[w,g]=0$ and $g\rho= \rho$. The particular form of $\tilde w$ implies that $\tr_B (\tilde{w} \rho \tilde{w}')= \tr_B(\tilde{w}' \tilde{w} \rho)=0$ unless $\tilde{w}'= \tilde{w}^{-1}$. This follows from $\tr_B(\tilde{w}\rho)= \sum_{gg'\in G} g_A \ket{0_A} \bra{0_A} g_A' g_A^{-1}\, \bra{0_B} g_B' \tilde{w} \ket{0_B}$ and for $\tilde{w} \neq \bbbone$ the scalar product $\bra{0_B} g_B' \tilde{w} \ket{0_B}$ is always zero. The reduced density matrix becomes
\begin{eqnarray}
\nonumber
\rho_{A}^{phys} &=&  \sum_{w,w' \in \mathcal W} c(w')\overline{c}(w^{-1}) \tr_B (\tilde{w}'\rho \tilde{w})\\
\nonumber
&=& \sum_{\tilde{w},\tilde{w}' \in \mathcal W} c(\tilde{w}') \overline{c}(\tilde{w}^{-1}) \delta_{\tilde{w}'\tilde{w}^{-1}} \tr_B \rho\\
&=& \rho_A
\end{eqnarray}
Then the von Neumann entropy for every physical state is equal to the entropy of the $G$-homogeneous state $\rho$.  \qed

\noindent {\em Entropic area law.} The entropy of the physical states in $\h_{phys}$ satisfies the {\em area law}. Consider a partition $(A,B)$ of $\h_{phys}$ constructed by taking as subsystem $A$ all the degrees of freedom (i.e., the local Hilbert spaces $\CC^{d_i}$ corresponding to the edges of the graph) inside or intersected by a closed surface $\Sigma$. The group $G_{AB}$ will be generated by all the stars $A_k$ based on sites outside the surface that puncture the surface $\Sigma$. Let $n_{AB}$ the number of such stars. Then the entropy is $S=\log_2|G_{AB}|=\log_2 f(n_{AB})$, where the function $f(n_{AB}):= |G_{AB}|$ gives the order of the group as a function of the number of its generators. As a measure of the surface $\Sigma$ we can choose the number of punctures $\sigma$. In general, $\sigma \ge n_{AB}$ because a star can puncture the surface in more than one point. It immediately follows that
\be
S\le \log_2 f(\sigma)
\label{arealaw}
\ee
This bound can be saturated for some geometries, e.g., in a cubic lattice if we choose $\Sigma$ to be convex \cite{hiz2}.

\section{$G$-homogeneous states for spin flips}\label{spinflips}

\subsection{Qubits}\label{qubits}

In our previous work \cite{hiz1,hiz2} we investigated examples of $G$-homogeneous states where $G$ was a group of spin flips and hence the states were also stabilizer states; in this case $G$ is Abelian and moreover $\forall g\in G,\ g^2=\bbbone$. We will clarify later the relationship between this class of $G$-homogeneous states and the well-known graph states. More exactly, we will prove that $G$-homogeneous states corresponding to a group of spin flips are locally equivalent to 2-colorable (i.e., bipartite) graph states. This representation of 2-colorable graph states as an equal superposition of elements of $G$ acting on a reference state proves to be very effective in calculating entanglement and general correlation functions.

In order to make the connection between $G$-homogeneous states and stabilizer states we start with some general considerations about the stabilizer formalism. Consider a system of $n$ spins-1/2 with Hilbert space $\mathcal H= \mathcal{H}_1^{\otimes n}$ ($\mathcal{H}_1= \mbox{span} \{ \ket{0}, \ket{1}\}$ is the Hilbert space of a single spin). As before, we choose the reference state $\ket{0}:= \ket{0}_1 \otimes \ldots \otimes\ket{0}_n$, corresponding to the basis vector with all spins up. We denote the Pauli matrices by $X_i$, $Y_i$ and $Z_i$ (the subscript represents the qubit on which they act). There are two necessary and sufficient conditions for a group of Pauli operators ${\cal S}= \langle s_1,\ldots, s_n \rangle$ to be a stabilizer group \cite{nielsen_chuang, gottesman}: (i) $s_i^2=\bbbone,\ s_i\ne -\bbbone,\ \forall i$; and (ii) $[s_i, s_j]= 0, \ \forall i,j$. From the general theory we know that any element of a stabilizer group can be written as $\pm X({\bf a}) Z({\bf b})$, where ${\bf a,b}\in \bbbz_2^n$ are binary vectors and $X({\bf a}):= \prod_{k=1}^n X_k^{a_k}$; analogously, $Z({\bf b})= \prod_{k=1}^n Z_k^{b_k}$ \cite{ortho_geometry}. We can define the $n\times 2n$ generator matrix as follows
\be
\begin{pmatrix}
{\bf a}_1 & \vline & {\bf b}_1 \cr
\vdots & \vline & \vdots \cr
{\bf a}_n & \vline & {\bf b}_n \cr
\end{pmatrix}
\ee
where $s_i= \pm X({\bf a}_i) Z({\bf b}_i)$ and ${\bf a}_i, {\bf b}_i\in \bbbz_2^n$ are the binary vectors corresponding to the generators (the sign is omitted in the definition of the generator matrix). Hence the left (right) part of the generator matrix contains the $X(Z)$-type generators, respectively. We also define the inner product of two binary vectors as
\be
({\bf a}, {\bf b}):= \sum_{i=1}^n a_i b_i \mod 2
\ee
The two conditions satisfied by the generators become: 
\begin{eqnarray}
&(i)& \ ({\bf a}_i, {\bf b}_i)= 0 \\
&(ii)& \ ({\bf a}_i, {\bf b}_j)+ ({\bf a}_j, {\bf b}_i)=0 \mod 2,\ \forall i,j
\label{generators}
\end{eqnarray}
An $n$-qubit {\em stabilizer state} $\ket{\psi}$ is stabilized by a group ${\cal S}_{\ket{\psi}}$ of Pauli operators having $n$ generators.

For example, the Bell states $\ket{\Phi^\pm}= 2^{-1/2}(\ket{00}\pm \ket{11})$ and $\ket{\Psi^\pm}= 2^{-1/2}(\ket{01}\pm \ket{10})$ are stabilized by the groups $\langle \pm X_1 X_2, Z_1 Z_2\rangle$ and $\langle \pm X_1 X_2, -Z_1 Z_2\rangle$, respectively and the generator matrix is $\begin{pmatrix} 11 & \vline & 00 \cr 00 & \vline & 11 \end{pmatrix}$.

Let ${\mathcal C}_1= \langle H, P \rangle$ be the 1-qubit Clifford group, mapping products of Pauli matrices into products of Pauli matrices; $H= 2^{-1/2}(X+Z)$ is the Hadamard and $P= \mbox{diag}(1, i)$.

A natural question to ask is when two stabilizer states are locally equivalent and hence they have the same entropy $S$. A restricted criterion for local Clifford equivalence only is the following:

\noindent {\em Proposition 3.} Let $G$ and $G'$ be two stabilizer groups of $n$ qubits and let $G$ be generated by $G= \langle g_1,...,g_n \rangle$. Then the states stabilized by $G$ and $G'$ are locally Clifford equivalent iff there is a local unitary $U\in {\mathcal C}_1^{\otimes n}$ such that the set $h_i= U g_iU^\dagger ,\ i=1,...,n$ generates $G'= \langle h_1,...,h_n \rangle$.

\noindent {\em Proof.} Denote by $\rho_G$ and $\rho_{G'}$ the states stabilized by $G$ and $G'$, respectively. Since $G$ is generated by $\langle g_1,...,g_n \rangle$, we can write
\begin{eqnarray}
\rho_G&=& \ket{G}\bra{G}= 2^{-n} \prod_{i=1}^n (1+g_i)
\end{eqnarray}
The two stabilizer states are locally Clifford equivalent iff there is a local unitary $U\in {\mathcal C}_1^{\otimes n}$ such that $\rho_{G'}= U\rho_G U^\dagger= 2^{-n} \prod_{i=1}^n (1+ U g_i U^\dagger)$; define $h_i:= U g_iU^\dagger ,\  i=1,...,n$. Then $h_i^2=\bbbone$ and $[h_i, h_j]=0,\ \forall i,j$, since the generators $g_i$ satisfy the same relations. Then $h_i$ can be chosen as a set of generators for $G'$. \qed

Turning to the previous example, it is easy to see that $\ket{\Phi^-}= Z_1\ket{\Phi^+}$ and hence their generators are related by $Z_1 (X_1 X_2) Z_1=- X_1 X_2$, $Z_1 (Z_1 Z_2) Z_1= Z_1 Z_2$.

Let $G\subseteq \langle X_1,\ldots, X_n \rangle$ be a group of spin flips: $\forall g\in G, g= X({\bf a})$, where ${\bf a}\in \bbbz_2^n$ is a binary vector and $X({\bf a})= \prod_{i=1}^n X_i^{a_i}$. Consider the generators of $G= \langle g_1,\ldots, g_k \rangle= \langle X({\bf a}_1),\ldots, X({\bf a}_k) \rangle$; $|G|= 2^k$. Then we can write the group $G$ in terms of an (additive) group of binary vectors ${\cal A}= \langle {\bf a}_1,\ldots, {\bf a}_k\rangle \subset \bbbz_2^n$, and therefore $G= X({\cal A})$.

For a group $G$ of spin flips, the $G$-homogeneous state is:
\be
\ket{G}= |G|^{-1/2} \sum_{{\bf a}\in {\cal A}} X({\bf a)} \ket{0}
\label{G0}
\ee
As mentioned before, $G$ leaves invariant the state $\ket{G}$, since $g \ket{G}= \ket{G}, \forall g\in G$. We now construct the stabilizer of $\ket{G}$, ${\cal S}_{\ket{G}}= \langle g_1,..,g_k, s_{k+1},.., s_n\rangle$; obviously it should have $n$ generators. The first $k$ are the pure spin flips generating $G$, $g_i= X({\bf a}_i), 1\le i \le k$; the other generators can be written as $s_j= \pm X({\bf a}_j) Z({\bf b}_j)$ with $k+1\le j\le n$. Since $s_j \ket{G}=\ket{G}$ and $[s_j, g]=0,\ \forall g$, we have
\begin{eqnarray}
\nonumber
1&=& \bra{G} s_j \ket{G}= \pm |G|^{-1} \sum_{g,g'\in G} \bra{0} g X({\bf a}_j) Z({\bf b}_j) g'\ket{0} \\
&=& \pm \sum_{h\in G} \bra{0} h X({\bf a}_j) \ket{0}= \pm \chi_G(X({\bf a}_j))
\end{eqnarray}
where the characteristic function is defined as $\chi_G(h)= 1$ if $h \in G$ and 0 otherwise. This implies that $\pm X({\bf a}_j)\in G$, hence we can choose all the generators $s_j$ as pure phase flips, $s_j= Z({\bf b}_j)$. Therefore the stabilizer ${\cal S}_{\ket{G}}= \langle X({\bf a}_1),..,X({\bf a}_k), Z({\bf b}_{k+1}),..,Z({\bf b}_n) \rangle= X({\cal A})\cdot Z({\cal B})$ is a direct product of pure spin flips and pure phase flips, with ${\cal A}= \langle {\bf a}_1,..,{\bf a}_k \rangle$ and ${\cal B}= \langle {\bf b}_{k+1},..,{\bf b}_n \rangle$. Commutation of all generators implies $({\bf a}_i, {\bf b}_j)=0,\ \forall i,j$, hence ${\cal B}= {\cal A}^\perp$. Thus the whole information about the state is contained only in the $X$-type generators (or, equivalently, only in the $Z$-type ones):
\be
{\cal S}_{\ket{G}}= X({\cal A})\cdot Z({\cal A}^\perp)
\ee
and so the group $G= X({\cal A})$ describes fully the stabilizer state $\ket{G}$. A stabilizer state having only $X$- and $Z$-type generators is also known as a Calderbank-Shor-Steane (CSS) state.

Another way of seeing this is the following. The stabilizer of the vacuum is ${\cal S}_{\ket{0}}:= \langle Z_1, \ldots, Z_n \rangle$ and hence:
\be
\ket{0}\bra{0}= 2^{-n} \sum_{h \in {\cal S}_0} h = 2^{-n} \sum_{{\bf b}\in \bbbz_2^n} Z({\bf b})
\ee
As any group of pure phase flips $Z({\cal B})\subseteq {\cal S}_{\ket{0}}$, these symmetries are already included in the way we constructed the $G$-homogeneous state (\ref{G0}), since the reference state is exactly $\ket{0}$.

\subsection{Relationship to graph states}\label{graph}

Graph states are a class of multiparticle entangled states which include cluster states, GHZ states etc and have been extensively studied recently \cite{graph1,graph2,graph3}. Given a graph $\mathcal G$, the associated graph state is stabilized by the following $n$ generators:
\be
g_i= X_i \prod_{j\in neigh(i)} Z_j,\ \ i=1,n
\ee
where the product is taken over all the nearest-neighbor vertices of vertex $i$. It is easy to see that the $n\times 2n$ matrix describing the generators is the following:
\be
(\bbbone_n | M )
\ee
hence the $X$-part of the generators is the unit matrix and the $Z$-part the adjacency matrix $M$ of the graph. Let us analyze first the relationship between stabilizer states and graph states. Clearly, graph states are a particular class of stabilizer states. According to Gottesman (\cite{gottesman}, Chapter 4) we can put the stabilizer matrix into a standard form by performing Gaussian elimination. For stabilizer states (set $k=0$ in Eq.(4.3) of \cite{gottesman}) we have the standard form
\be
\begin{pmatrix}
\bbbone_{r, r} & A_{r, n-r} & \vline & B_{r, r} & C_{r, n-r} \cr 
0_{n-r, r} & 0_{n-r, n-r} & \vline & D_{n-r, r} & \bbbone_{n-r, n-r}
\end{pmatrix}
\ee
where $r$ is the rank of the $X$ part of the generators matrix (the subscripts denote the size of the matrices). If $r=n$, then this is equal to
\be
\begin{pmatrix}
\bbbone\ \vline\ B \cr 
\end{pmatrix}
\ee
with both $n\times n$-matrices. Now, is $B$ the adjacency matrix of a graph? We need to prove that $B_{ii}=0$ and $B_{ij}=B_{ji}$. This follows immediately from the properties of stabilizer generators. For any generators $g_{i,j}= X({\bf a}_{i,j}) Z({\bf b}_{i,j})$, we have $({\bf a}_i, {\bf b}_i)= 0$ and $({\bf a}_i, {\bf b}_j)+ ({\bf a}_j, {\bf b}_i)=0 \mod 2,\ \forall i,j$, from which follows that $B$ is indeed an adjacency matrix. Therefore, if the rank of the $X$ part of the generator matrix is maximal, $r=n$, the stabilizer state in question is a graph state. What if $r<n$? It has been shown that any stabilizer state is equivalent to a graph state under local Clifford operation \cite{nest}. In conclusion, an arbitrary stabilizer state is either a graph state (if $r=n$), or is locally equivalent to a graph state ($r<n$).

One problem with graph states is that there is no 1-to-1 correspondence between graphs and locally inequivalent states. Thus, a GHZ state can be described either by a star graph or by a fully connected graph \cite{bell_graph}. Moreover, some graph states have a more compact description as $G$-homogeneous states. For example, the stabilizer of the GHZ state is 
\be
{\cal S}_{GHZ}= \langle X^{\otimes n}, Z_1Z_2,..., Z_1 Z_n \rangle
\ee
with the notation $X^{\otimes n}:= \prod_{i=1}^n X_i$. Described as a graph state, the GHZ state is (locally equivalent) to a star graph with a stabilizer group generated by 
\be
{\cal S'}_{star}= \langle X_1 \prod_{i=2}^n Z_i, X_2 Z_1,..., X_n Z_1 \rangle
\ee
It is immediate to see that the two set of generators are related by $g'_k= {\bf H} g_k {\bf H},\ k=1,...,n$, where the local operator is ${\bf H}:= \prod_{k=2}^n H_k$ ($H_k$ is a Hadamard on the $k$ qubit). Then the density matrix of the star graph state stabilized by ${\cal S'}_{star}$ is
\be
\rho_{star}= 2^{-n} (1+ X_1 \prod_{i=2}^n Z_i) \prod_{j=2}^n (1+ X_j Z_1)
\ee
The above form is more complicated than the description of $\ket{GHZ}$ in our formalism, since the group $G$ of spin flips is just $G= \langle X^{\otimes n} \rangle$, $|G|= 2$. So instead of describing the $n$-GHZ state using $n$ generators, we use only one, irrespectively of the number of qubits. Thus the rationale behind this approach is to find a more convenient description for a certain class of stabilizer/graph states.

In the last section we proved that the generators matrix associated to the $G$-homogeneous state is:
\be
\begin{pmatrix}
A_{k, n} & \vline & 0_{k, n} \cr
0_{n-k, n} & \vline & A_{n-k, n}^\perp 
\end{pmatrix}
\label{hiz_gen}
\ee
where $0_{mn}$ is the $m\times n$ zero matrix and $A= ( {\bf a}_1,..,{\bf a}_k )^T$. Intuitively, if we can separate the $X$- and $Z$-type generators, we can ``throw away'' the $Z$ part and work only with the spin flips (acting on vacuum), since the $Z$'s leave invariant the vacuum $\ket{0}$.

Consider now some transformations. A Hadamard on the $i$-qubit $H_i$ interchanges $Z_i \leftrightarrow X_i$, and it is equivalent to exchanging the $i$-column in the $X$ part with the $i$ column in the $Z$ part of the generator matrix (\ref{hiz_gen}). Therefore applying $H^{\otimes n}$ interchanges the $X$ and $Z$ blocks and this is convenient if $n-k<k$, since we have a local equivalent state described by a smaller group of spin-flips. The next proposition establishes the relationship between graph states and $G$-homogeneous states.

\noindent {\em Proposition 4.} Let $\mathcal G$ be a bipartite graph (i.e., 2-colorable) and let $V_1, V_2$, with $|V_1|\le |V_2|$, be the sets of vertices forming the bipartition (thus all the vertices in $V_i$ have the same colour). Then the associated graph state $\ket{\mathcal{G}}$ is locally equivalent to a $G$-homogeneous state, $\ket{G}= U\ket{\mathcal{G}}$, with $U= \prod_{i\in V_2} H_i$. Moreover, the group $G$ of spin-flips satisfies $|G|= 2^{|V_1|}$.

\noindent {\em Proof.} Without loss of generality, we can label the vertices such that the first $n_1:= |V_1|$ belong to the $V_1$ partition; define also $n_2:= |V_2|= n-n_1$. Then the generator matrix of the stabilizer is
\be
\begin{pmatrix}
\bbbone_{n_1} & 0_{n_1,n_2} & \vline & 0_{n_1} & A_{n_1,n_2} \cr
0_{n_2,n_1} & \bbbone_{n_2} & \vline & A_{n_2,n_1}^T & 0_{n_2} \cr
\end{pmatrix}
\ee
where the subscripts denote the size of the matrices (for square matrices only one index is used); $A$ is the non-zero part of the adjacency matrix. Performing a Hadamard on all the qubits belonging to $V_2$ interchanges the $X$ and $Z$ columns:
\be
\begin{pmatrix}
\bbbone_{n_1} & A_{n_1,n_2} & \vline & 0_{n_1} & 0_{n_1,n_2} \cr
0_{n_2,n_1} & 0_{n_2} & \vline & A_{n_2,n_1}^T & \bbbone_{n_2} \cr
\end{pmatrix}
\ee
which is of the form (\ref{hiz_gen}), so we can write
\be
\ket{G}= \prod_{i\in V_2} H_i \ket{{\mathcal G}}
\ee
We can also check that all generators commute, hence $[X({\bf a}_i), Z({\bf b}_j)]=0$, $\forall i=1,...,n_1$ and $j=1,...,n_2$. We have $({\bf a}_i, {\bf b}_j)= A_{ji}^T+ A_{ij}= 0\mod 2$, as expected (this should have been obvious, since applying local Hadamards to a generators matrix gives another generators matrix). Since the number of $X$ generators is $n_1$, we have $|G|= 2^{|V_1|}$. \qed

This result has been independently proved by Chen and Lo \cite{lo}. Moreover, they also proved the reciprocal: any CSS state is locally equivalent to a 2-colorable graph state. Therefore we conclude that if $G$ is a group of spin flips acting on qubits, the corresponding $G$-homogeneous states are locally equivalent to 2-colorable graph states.

Applying Eq.(\ref{S}), it is now easy to derive a bound on the entropy for bipartite graph states:
\be
S\le \log_2 |G|= |V_1| \le \left\lfloor \frac n 2 \right\rfloor
\ee
recovering a result obtained in \cite{graph1}.

Examples of 2-colorable graphs include cluster states, trees ($n$-star, $n$-linear) and $2n$-ring graphs. Since they are equivalent to $G$-homogeneous states, they have a simpler description in terms of only $X$-type generators. How widespread are 2-colorable graph states? In Ref.~\cite{graph1} the authors found 45 (connected) graph states with up to seven vertices which are not equivalent (under local unitaries and graph isomorphisms); out of these, 32 are 2-colorable and the other 13 are 3-colorable.

\subsection{Qudits}\label{qudits}

The formalism of $G$-homogeneous states can also encompass multiqudit states. A {\em qudit} is a quantum system having a $d$-dimensional Hilbert space $\h_d= \mbox{span} \{ \ket{0},\ldots, \ket{d-1} \}\cong \CC^d$. The generalized Pauli operators for qudits are defined as
\begin{eqnarray}
X \ket{k}&=& \ket{k\oplus 1} \\
Z \ket{k}&=& \omega^k\ \ket{k}
\end{eqnarray}
where $\omega= e^{2\pi i/d}$ and $\oplus$ is the sum modulo $d$. They are no longer idempotent, since $X^d= Z^d= \bbbone$; moreover $ZX= \omega\, XZ$. It is straightforward to see that for the case of qubits ($d=2$) we recover the usual definitions and commutations relations for $X$ and $Z$. We can generate any basis vector by applying spin flips on $\ket{0}$, $\ket{k}= X^k \ket{0}$, hence an arbitrary state in $\h_d$ has the form $\ket{\psi}= \sum_{k=0}^{d-1} \alpha_k X^k \ket{0}$.

As an example, consider the Hilbert space of $n$ qudits $\h= (\CC^d)^{\otimes n}$ and let $G= \langle X^{\otimes n} \rangle= \{ \bbbone, X^{\otimes n},...,(X^{\otimes n})^{d-1} \}$, with $X^{\otimes n}= \prod_{i=1}^n X_i$; obviously $|G|= d$. The corresponding $n$-qudit $G$-homogeneous state is
\be
\ket{G}= d^{-1/2}\sum_{i=0}^{d-1} (X^{\otimes n})^i \ket{0}
\ee
It is immediate to see that $\ket{G}$ is the maximally entangled state of $n$ qudits (it generalizes the $n$-GHZ state), since $d_A=d_B=1$ for {\em any} bipartition $(A,B)$, so the entropy is $S= \log_2 d$ as expected.

\section{Multiparticle entanglement}\label{ntangle}

In this section we show that the compact form of writing a $G$-homogeneous state (\ref{g_state}) is a useful device in calculating multiparticle entanglement. The von Neumann entropy characterizes well the {\em bipartite} entanglement of a system in a pure state. However, $S$ fails to address the problem of multipartite entanglement. This becomes apparent even for the simple case of three qubits; there are two classes of states with genuine tripartite entanglement, the $W$- and $GHZ$-type states, which cannot be distinguished by calculating only the bipartite entanglement \cite{wstate}. An entanglement measure which does distinguish between the two families of states is the {\em 3-tangle} $\tau_3$ \cite{3tangle}, since $\tau_3(GHZ)=1$ and $\tau_3(W)=0$. More generally, $\tau_3(\psi_{GHZ})> 0$, and $\tau_3(\psi_W)=0$ for all the states $\psi_{GHZ}, \psi_W$ belonging to the $GHZ$ and the $W$ family, respectively.

A generalization of the 3-tangle to an even number $n$ of qubits is the {\em n-tangle} introduced in Ref.~\cite{ntangle}. For a state $\ket{\psi}$, the $n$-tangle is defined as:
\be
\tau_n:= |\bra{\psi} Y^{\otimes n} \ket{\psi^*} |^2 
\ee
where $*$ means complex conjugation. It has been shown that $\tau_n$ is an entanglement monotone and is invariant under local unitaries. Note that $\tau_n$ is not defined for $n$ odd.

\noindent {\em Proposition 5.} Let $G= \langle X({\bf a}_1),..,X({\bf a}_k) \rangle$ be a group of spin flips. Denote by $p_i:= ({\bf a}_i, {\bf a}_i)$ the parity of the generator $g_i= X({\bf a}_i)$ and let $p(h)= ({\bf a}, {\bf a})$ be the parity of an arbitrary element $h= X({\bf a})\in G$. Then we have the following
\be
|G|^{-1} \sum_{h \in G} (-1)^{p(h)}= \prod_{i=1}^k (1-p_i) =: 1-p(G)
\ee
where the last equation defines the parity $p(G)$ of the group $G$. The proof is trivial. If all the generators are even, then all the group elements have even parity, hence the above sum is 1 and $p(G)=0$. If only one of the generators is odd (say $g_1$), then exactly half of the group elements are odd (those containing $g_1$ in their expansion) and the other half are even; therefore the sum is 0 and $p(G)=1$. If more than one generator is odd, we can always choose an equivalent set of generators such that only one is odd, say $g_1$, e.g., by multiplying all the odd generators by $g_1$ (apart from $g_1$ itself); hence this case reduces to the previous one.

We now calculate $\tau_n$ for $G$-homogeneous states (and, consequently, for 2-colorable graph states, as they are locally equivalent). Let $\ket{\psi}= |G|^{-1/2} \sum_{g\in G} g \ket{0}= |G|^{-1/2} \sum_{{\bf a}\in A} X({\bf a)} \ket{0}= \ket{\psi^*}$. Since $Y^{\otimes n} X({\bf b})= (-1)^{({\bf b}, {\bf b})} X({\bf b}) Y^{\otimes n}= i^n (-1)^{({\bf b}, {\bf b})} X({\bf b}) X^{\otimes n} Z^{\otimes n}$ and $Z^{\otimes n} \ket{0}= \ket{0}$, we obtain
\begin{eqnarray}
\nonumber
\bra{G}Y^{\otimes n}\ket{G} &=& |G|^{-1}\sum_{{\bf a,b}\in {\cal A}} \bra{0} X({\bf a}) Y^{\otimes n} X({\bf b}) \ket{0} \\
\nonumber
&=& i^n |G|^{-1} \sum_{{\bf a,b}\in {\cal A}} (-1)^{({\bf b}, {\bf b})} \bra{0} X({\bf a}) X({\bf b}) X^{\otimes n} \ket{0} \\
\nonumber
&=& i^n |G|^{-1} \sum_{{\bf b,b'}\in {\cal A}} (-1)^{({\bf b}, {\bf b})} \bra{0} X({\bf b'}) X^{\otimes n} \ket{0} \\
&=& i^n (1-p(G))\ \chi_G(X^{\otimes n})
\end{eqnarray}
where the characteristic function $\chi_G(X^{\otimes n})= 1$ if $X^{\otimes n}\in G$ and 0 otherwise. Therefore we have the following:\\
{\em Corollary.} For a $G$-homogeneous state the $n$-tangle is
\be
\tau_n= (1-p(G))\ \chi_G(X^{\otimes n})
\ee
Hence the $n$-tangle is 1 iff all the generators of $G$ are even and $X^{\otimes n}\in G$.

\section{Summary}\label{summary}

Entanglement has been used recently to probe many body systems and gain insights of their properties. Quantum phase transitions are a notable example in this sense. Several authors have shown that a critical behavior of entanglement can signal quantum phase transitions \cite{phasetrans}.

However, calculating the entanglement entropy for an arbitrary system is often a computationally intractable problem. One way to circumvent this problem is to focus on states with extra built-in symmetries. In the present article we formalized this intuition using tools from group theory. We have investigated the entanglement entropy for a class of states constructed by acting with the group algebra of a possibly non-Abelian group $G$ on a separable reference state $\ket{0}$. The group is required to have a bilocal action with respect to a given partition $A,B$ of the full Hilbert space $\h= \h_A\otimes \h_B$. We started first with the $G$-states, which are constructed as an arbitrary superposition of group elements acting on $\ket{0}$. These states are extremely general, as any state of a Hilbert space can be regarded as a $G$-state with an suitable group $G$. Moreover, we have shown that the ground states of generic Hamiltonians are $G$-states. We have derived an upper bound for the entropy provided a separability condition holds for the coefficients.

A particular class are the $G$-homogeneous states and in this case we generalized our previous results \cite{hiz1,hiz2}. If $G$ is a group of spin flips, we show that the associated $G$-homogeneous states are locally equivalent to 2-colorable graph states and CSS states. Examples include GHZ states and the ground state of the Kitaev model.

We have shown that we can regard the $G$-states as the physical states in a quantum gauge theory. In this framework all the physical states are obtained acting on the $G$-homogeneous state with the commutant algebra of the group algebra of $G$. With some extra assumptions, we can compute the bipartite entanglement for all the physical states. We have shown how to relate this quantity to the geometric entropy introduced in \cite{callan}, namely the von Neumann entropy relative to a bipartition obtained by considering a closed surface $\Sigma$ of area $\sigma$ and taking as subsystem $A$ all the particles (or degrees of freedom) within $\Sigma$. Moreover, the entanglement of the physical states obeys the area law, i.e., $S(\ket{\phi_{phys}})= f(\sigma)$.

Finally, we have shown that this construction can be extended in two directions: computing the entanglement entropy for qudits in $G$-homogeneous states and the $n$-tangle for 2-colorable graph states.

A future challenge will be to find other physical systems which can benefit from the group theoretical framework described here.



\begin{thebibliography}{}

\bibitem{nielsen_chuang} M.A.~Nielsen and I.L.~Chuang, {\em Quantum computation and quantum information} (Cambridge University Press, Cambridge, 2000).

\bibitem{zanardi} P.~Zanardi, X.~Wang, J.~Phys.~A {\bf 35}, 7947 (2002); quant-ph/0201028.

\bibitem{botero} A.~Botero, B.~Reznik, {\em BCS-like Modewise Entanglement of Fermion Gaussian States}, quant-ph/0404176.

\bibitem{fqhe} X.G.~Wen, \pl A {\bf 300}, 175 (2002).

\bibitem{phasetrans} T.J.~Osborne and M.A.~Nielsen, \pra {\bf 66}, 32110 (2002); quant-ph/0202162; A.~Osterloh, L.~Amico, G.~Falci, and R.~Fazio, Nature {\bf 416}, 608 (2002); quant-ph/0202029; J.~Vidal, G.~Palacios, and R.~Mosseri, \pra {\bf 69}, 22107 (2004); quant-ph/0305573; Y.~Chen, P.~Zanardi, Z.D.~Wang, and F.C.~Zhang, {\em Entanglement and Quantum Phase Transition in Low Dimensional Spin Systems}, quant-ph/0407228.

\bibitem{cardy} P.~Calabrese, J.~Cardy, J.~Stat.~Mech.~{\bf 2004}, P06002 (2004);  hep-th/0405152.

\bibitem{cirac} F.~Verstraete, M.~Popp, and J.~I.~Cirac, \prl {\bf 92}, 27901 (2004); quant-ph/0307009; F.~Verstraete, M.A.~Martin-Delgado, J.I.~Cirac, \prl {\bf 92}, 087201 (2004); quant-ph/0311087.

\bibitem{keating} J.P.~Keating and F.~Mezzadri, \prl {\bf 94}, 050501 (2005); J.P.~Keating and F.~Mezzadri, Comm.~Math.~Phys.~{\bf 252}, 543 (2004).

\bibitem{latorre} G.~Vidal, J.~I.~Latorre, E.~Rico, and A.~Kitaev, \prl {\bf 90}, 227902 (2003); quant-ph/0211074; J.~I.~Latorre, E.~Rico, and G.~Vidal, Quant.~Inf.~and Comp.~{\bf 4}, 48 (2004); quant-ph/0304098.

\bibitem{korepin} A.R.~Its, B.Q.~Jin, V.E.~Korepin, J.Phys. A {\bf 38}, 2975 (2005); quant-ph/0409027.

\bibitem{plenio_vedral} M.B.~Plenio and V.~Vedral, Contemp.Phys. {\bf 39}, 431-466 (1998).

\bibitem{hp} G.~t' Hooft, {\em Dimensional Reduction in Quantum Gravity}, gr-qc/9310026; L.~Susskind, J.~Math.~Phys.~{\bf 36}, 6377 (1995); R.~Bousso, \rmp {\bf 74}, 825 (2002).

\bibitem{srednicki} M.~Srednicki, \prl {\bf 71}, 666 (1993).

\bibitem{plenio} M.B.~Plenio, J.~Eisert, J.~Drei\ss ig, and M.~Cramer, \prl {\bf 94}, 060503 (2005); quant-ph/0405142.

\bibitem{wolf} M.M.~Wolf, {\em Entropic area law for fermions}, quant-ph/0503219.

\bibitem{css} A.R.~Calderbank and P.W.~Shor, \pra {\bf 54}, 1098 (1996); A.M.~Steane, \prl {\bf 77}, 793 (1996).

\bibitem{kitaev} A.Y.~Kitaev, Ann.~Phys.~(N.Y.) {\bf 303}, 2 (2003); quant-ph/9707021.

\bibitem{fattal} D.~Fattal, T.S.~Cubitt, Y.~Yamamoto, S.~Bravyi, and I.L.~Chuang, {\em Entanglement in the stabilizer formalism}, quant-ph/0406168.

\bibitem{hiz1} A.~Hamma, R.~Ionicioiu, and P.~Zanardi, Phys.~Lett.~A {\bf 337}, 22 (2005); quant-ph/0406202.

\bibitem{hiz2} A.~Hamma, R.~Ionicioiu, and P.~Zanardi, \pra {\bf 71}, 022315 (2005); quant-ph/0409073.

\bibitem{dirac} P.A.M.~Dirac, {\em Lectures in Quantum Mechanics} (Dover Publications, New York, 2001).

\bibitem{ortho_geometry} A.R.~Calderbank, E.M.~Rains, P.W.~Shor, and N.J.A.~Sloane, \prl {\bf 78}, 405 (1997).

\bibitem{graph1} M~Hein, J.~Eisert and H.J.~Briegel, \pra {\bf 69}, 062311 (2004).

\bibitem{graph2} H.~Aschauer, W.~Dur and H.J.~Briegel, \pra {\bf 71}, 012319 (2005).

\bibitem{graph3} R.~Raussendorf, D.E.~Browne and H.J.~Briegel, \pra {\bf 68}, 022312 (2003).

\bibitem{gottesman} D.~Gottesman, {\em Stabilizer Codes and Quantum Error Correction} (Caltech Ph.D.~Thesis), quant-ph/9705052.

\bibitem{nest} M.~Van den Nest, J.~Dehaene, and B.~De Moor, \pra {\bf 69}, 022316 (2004); quant-ph/0308151.

\bibitem{bell_graph} O.~Guehne, G.~Toth, P.~Hyllus, H.J.~Briegel, {\em Bell Inequalities for Graph States}, quant-ph/0410059.

\bibitem{lo} K.~Chen, H.K.~Lo, {\em Conference Key Agreement and Quantum Sharing of Classical Secrets with Noisy GHZ States}, quant-ph/0404133.

\bibitem{wstate} W.~Dur, G.~Vidal, and J.I.~Cirac, \pra {\bf 62}, 062314 (2000); quant-ph/0005115.

\bibitem{3tangle} V.~Coffman, J.~Kundu, and W.K.~Wootters, \pra {\bf 61}, 052306 (2000); quant-ph/9907047.

\bibitem{ntangle} A.~Wong and N.~Christensen, \pra {\bf 63}, 044301 (2001); quant-ph/0010052.

\bibitem{callan} C.~Callan and F.~Wilczek, \pl {\bf B 333}, 55 (1994); G.~'t Hooft, Nucl.Phys.B {\bf 256}, 727 (1985).

\end{thebibliography}
\end{document}